A coupling of martensitic and metamagnetic transitions with collective magneto-volume and table-like magnetocaloric effects

E. K. Liu,[1] Z. Y. Wei,[1] Y. Li,[1,2] G. D. Liu,[2] H. Z. Luo,[2] W. H. Wang,[1] H. W. Zhang,[1] and G. H. Wu[1,a)]

[1]*State Key Laboratory for Magnetism, Beijing National Laboratory for Condensed Matter Physics, Institute of Physics, Chinese Academy of Sciences, Beijing 100190, China*

[2]*School of Material Science and Engineering, Hebei University of Technology, Tianjin 300130, China*

**Abstract**: A coupling of the first-order paramagnetic-to-induced-ferromagnetic martensitic and the second-order antiferromagnetic-to-ferromagnetic metamagnetic transitions was found in MnNi$_{0.8}$Fe$_{0.2}$Ge alloy. Based on the coupling, a magneto-volume effect driven by the martensitic transition and a table-like magnetocaloric effect generated by the successive magnetic phase transitions arise collectively. By using the magneto-volume effect, the internal pressure stress in the volume-expansion martensitic transition was determined at 350 MPa. The magnetocaloric effect, with a wide working temperature range of 26 K around room temperature, shows a small hysteresis loss (5 J/kg) and a large net refrigerant capacity (157 J/kg).

[a)] Author to whom correspondence should be addressed. Electronic mail: ghwu@iphy.ac.cn

The ferromagnetic martensitic transition (FM-MT) has been playing a key role in multifunctional ferromagnetic shape memory alloys (FSMAs).[1-5] Among the various FSMAs, the MTs are always accompanied by different changes of magnetic states, such as FM-to-FM,[6,7] FM-to-Paramagnetic (PM),[8] FM-to-Antiferromagnetic (AFM)[2,9] and PM-to-FM[5,10] changes. Based on these magnetostructural transitions, diverse physical effects[11-14] including the giant magnetocaloric effect (MCE),[15,16] have been extensively reported. Especially, it has been known that in Heusler $Ni_2MnGa$-based FSMAs the temperature ($T_t$) of MT and Curie temperature of austenite phase ($T_C^A$) can merge together by tuning Ni/Mn proportion[17] or substituting Cu for Mn[18], resulting in a coupling of MT and FM transition and a resultant giant MCE. To date, the coupling of MT and AFM-to-FM metamagnetic transition has seldom been found in FSMAs. Owing to the metamagnetism, this kind of coupled transitions, compared with the ferromagnetism in $Ni_2MnGa$ alloys,[17,18] may show more physical behaviors.

In our previous work,[5] a Curie-temperature window, spanning from 300 to 210 K, was established in hexagonal $MnNi_{1-x}Fe_xGe$ MM'X alloys. Within the window, a stable magnetostructural coupling of the first-order MTs were realized for the magnetoresponsive effects. For this window, the upper boundary is just at around 300 K, where the MT ($T_t$) encounters the PM-to-FM transition ($T_C^M$) of the martensite phase, with $T_t$ (~ 300 K) being slightly higher than $T_C^M$ (~ 291 K) (also see inset of Fig. 1). At this critical point, the corresponding alloy composition is $MnNi_{0.8}Fe_{0.2}Ge$. For this composition, the martensite phase just locates at the critical region of the AFM-to-FM transition due to the moderate Fe content that produces FM coupling against the native AFM matrix in $MnNi_{1-x}Fe_xGe$. In this letter, we report the lately-found critical behaviors in this alloy. A coupling of the MT and a metamagnetic transition results in successive magnetic phase transitions, in which a magneto-volume effect driven by the MT and a table-like magnetocaloric effect with a large net refrigerant capacity are collectively presented at room temperature.

Polycrystalline $MnNi_{0.8}Fe_{0.2}Ge$ was prepared by arc melting method. The ingots were annealed at 1123 K for five days and then slowly cooled to room temperature. The phase structures were characterized by powder x-ray diffraction (XRD) with Cu-*K*α radiation. The magnetization measurements were performed on a superconducting quantum interference device (SQUID).

XRD analysis of MnNi$_{0.8}$Fe$_{0.2}$Ge was performed at room temperature (298 K), as shown in Fig. 1. The austenite and martensite, indexed as Ni$_2$In-type hexagonal and TiNiSi-type orthorhombic structures respectively, coexist at room temperature. This is in good agreement with our previous study[5] that Fe substitution for Ni lowers $T_t$ of MnNiGe to 300 K. The comparable volume fractions of two phases indicate that only about 50% martensite phase is produced at 298 K. It is thus hopeful to expect that the subsequent transition of residual parent phase will couple with the PM-to-FM transition (~ 291 K) of martensite phase.[5] The calculated lattice parameters of both phases are listed in Table I. The cell volume increases remarkably by 3.52% during the hexagonal-to-orthorhombic MT. Note that the increase of $c_{hex}$ axis is as large as 12.04 %. This large expansion behavior is coherent with the cases in the MnNi(Co)Ge MM'X alloys.[5,19-22]

To further examine the coupling of the martensitic and magnetic transitions, we measured the thermomagnetization, M(T) curves of MnNi$_{0.8}$Fe$_{0.2}$Ge in different magnetic fields, as shown in Fig. 2(a). In low fields (for instance, 0.5 kOe), a typical PM-to-AFM transition at 291 K ($T_N^M$) is observed, with an AFM ground state below $T_N^M$ as well as very low magnetizations (less than 2 emu/g in the whole measured temperature range). This behavior is similar to the case of Fe-free MnNiGe compound, which has a spiral AFM structure in martensite phase.[5] Introducing Fe atoms into MnNi$_{1-x}$Fe$_x$Ge, the FM coupling was established based on the Fe-6Mn local configurations in martensite phase.[5] For MnNi$_{0.8}$Fe$_{0.2}$Ge, the moderate Fe content results in a competition between FM and the native AFM couplings, leaving an AFM ground state in martensite phase in low fields, as shown in Figs. 2(a) and 2(b). With increasing field, the competition balance is prone to be destroyed and the AFM ground state is changed to FM state (AFM-to-FM metamagnetic transition), leading to a fact that the Néel point ($T_N^M$) gradually becomes a Curie-like point ($T_C^M$ ~ 291 K). The martensite phase thus gains increasing magnetization in the induced FM state. Nevertheless, the AFM coupling becomes increasingly strong at low temperatures. As a result, the induced FM state in martensite phase again shows a tendency to retransform to the AFM state with a critical transition temperature of $T_{cr}$. This $T_{cr}$ decreases from $T_N^M$ with increasing magnetic field. Therefore, at the intermediate temperatures (not very far below Néel point), the magnetic field can much easily drive the AFM ground state to FM state via a metamagnetic transition in martensite phase, which will be further indicated by M(H) curves in Fig. 3.

In Fig. 2(a), one can further see that a sharp magnetic transition appears just above $T_N^M$ with a thermal hysteresis in high fields. A magnified image is given as Fig. 2(c). The clear thermal hysteresis suggests the first-order MT around 300 K, which is in line with the XRD analysis in Fig. 1. Since $T_N^M$ (~ 291 K) is very close to $T_t$ (~ 300 K) of MT, the short-range AFM interactions above $T_N^M$ are established in zero field in martensite phase once the martensite is produced during the MT. Like the AFM-to-FM transition below $T_N^M$, these short-range interactions are also induced to FM ones and are shifted to higher temperatures by high fields. These induced short-range FM interactions thus appear at the temperature $T_t$ where the martensite phase is produced. Therefore, the MT is characterized by a change from PM austenite phase to induced FM martensite phase, forming a PM-to-induced-FM MT. An observed shift of MT toward to high temperatures reveals the occurrence of magnetic-field-induced MTs. The corresponding region of PM-to-induced-FM MTs in different fields is marked as region 1 (also colored as light yellow). Following region 1, there is the region 2 (also colored as light green) of the above-mentioned AFM-to-FM metamagnetic transition in martensite phase below $T_N^M$.

From the above results, it can be seen that, in low fields the alloy undergoes upon cooling a MT from PM austenite to PM martensite and a magnetic transition from PM martensite to (spiral) AFM martensite; While in higher fields, the alloy undergoes a MT from PM austenite to induced-FM martensite and an AFM-to-FM metamagnetic transition in martensite, followed by a retransformation to the AFM martensite at low temperatures. The MT and the metamagnetic transition construct two successive magnetic phase transitions in this alloy. Note that many similar AFM-to-FM metamagnetic transitions have been found in rare-earth compounds.[23-25]

However, these transitions usually locate at very low temperatures. In present alloy, differently, the transition happens around room temperature and couples with a martensitic transition.

Based on the coupling of the martensitic and metamagnetic transitions, an interesting effect arises, as shown in Fig. 2(c). For the AFM ordering, the value of $T_N^M$ upon cooling is abnormally higher than that upon heating (3.3 K higher, see M(T) curves in fields of 0.5, 5 and 10 kOe and inset to Fig. 2(d)). For MnNiGe based alloys, a weak first-order nature of the AFM ordering at $T_N^M$, evidenced by a thermal hysteresis, has been previously confirmed due to the strong magneto-elastic coupling.[5,26] The stoichiometric MnNiGe shows a hysteresis of 4 K[5] and the Sn-substituted MnNiGe$_{0.98}$Sn$_{0.02}$ 5.4 K.[26] The AFM ordering of both alloys occurs without compressive stress since their MTs happen at higher temperatures. Assuming the hysteresis of MnNi$_{0.8}$Fe$_{0.2}$Ge is close to these values (average value of ~ 4.7 K), as displayed in the inset to Fig. 2(d), one can conclude that the $T_N^M$ is raised by ~ 8 K upon cooling relative to its intrinsic transition temperature without pressure (schematized by the dashed line). For the martensitic transitions with volume expansion, the volume change always produces compressive stress in the system. During the MT of MnNi$_{0.8}$Fe$_{0.2}$Ge, the volume change from hexagonal parent phase to orthorhombic martensite is as large as 3.52% (Fig. 1 and Table I), which will gain large stress and thus reduce the lattices of two phases. This behavior has been evidenced by the temperature-dependent XRD analysis in MnNiGe:Fe systems (see Fig. 2(d) of Ref. 5). On the other hand, in MnNi(Co)Ge MM'X alloys a magneto-volume effect has been previously studied.[27,28] A pressure dependence of $T_N^M$, showing a slope of $dT_N^M/dP = +$ 2.3 K/kbar (+ 0.023 K/MPa), was found in MnNiGe,[27] which means that $T_N^M$ increases with increasing stress. Thus, it can be deduced that in present alloy it is the stress produced during the volume-expansion MT that brings about a magneto-volume effect and consequently raises $T_N^M$ to high temperatures.

Referring to the pressure dependence of $T_N^M$ in MnNiGe,[27] one can estimate the compressive stress at ~350 MPa during the martensitic transitions in MnNi$_{0.8}$Fe$_{0.2}$Ge, corresponding to the 8-K increase of $T_N^M$. Upon heating, no stress is produced to affect $T_N^M$ (Figs. 2(c) and 2(d)) as the inverse MT shows a shrink of the cell volume. Therefore, a magneto-volume effect driven by the volume-expansion martensitic structural transition was found in MnNi$_{0.8}$Fe$_{0.2}$Ge based on the coupling of martensitic and AFM transitions. With the aid of this magneto-volume effect, the important parameter of internal stress in volume-expansion martensitic transitions can be obtained in this alloy.

In order to probe the magnetization behavior during the successive magnetic phase transitions, the isothermal M(H) curves were measured across the successive transitions, as shown in Fig. 3. The magnetization behavior can be divided to two styles between 320 and 225 K. Above 290 K, a small magnetic hysteresis can be seen between the magnetization and demagnetization curves at each temperature. This indicates the occurrence of magnetic-field-induced MTs during the magnetization process. Coherent with the case in Fig. 2(c), this transition region has also been marked as region 1. Below 290 K, in contrast, the small magnetic hysteresis in M(H) curves rapidly becomes zero. A metamagnetization at 275 K, with an S-shaped curve and a critical field ($H_{cr}$) of 5 kOe, is shown in the inset to Fig. 3. This corresponds to the second-order AFM-to-FM metamagnetic transition in martensite phase, which is in line with the case in Figs. 2(a), 2(b), 2(c). This transition region is also marked as region 2. In the region, a relatively large magnetization increment in the field range of 5 to 20 kOe can be seen due to the AFM-to-FM metamagnetic transition.

Using the isothermal M(H) curves, the magnetic-entropy changes ($\Delta S_m$) across the successive magnetic phase transitions were estimated according to the Maxwell relation,[29] as shown in Fig. 4(a). Notably, a table-like MCE is observed around room temperature. It can be seen that the $\Delta S_m$ plateau consists of peak 1 and 2 (two-peak structure), which corresponds to the PM-to-induced-FM MT and the AFM-to-FM metamagnetic transition, respectively, as shown in Figs. 2 and 3. One can thus see that the MCE in this alloy originates from two different contributions: one is the magnetostructural transition related to the spin-phonon coupling,[30] the other is the metamagnetic transition related to the magneto-elastic coupling.[31] At low fields, the peak 2 is higher than that of peak 1 due to the occurrence of the AFM-to-FM metamagnetic transition. In contrast, at high fields peak 1 increases quickly once the magnetic-field-induced MT begins, with a maximum of -8 J K$^{-1}$ kg$^{-1}$ for the field change ($\Delta H$) of 50 kOe. These field dependences of both $\Delta S_m$ peaks are further plotted in the inset to Fig. 4(a), showing an almost linear relation between $\Delta S_m$ and $\Delta H$. Based on the successive magnetic phase transitions, the full width at half maximum (FWHM) of the MCE plateau is expended to about 26 K. This range is wider than those of many first-order magnetostructural transitions with giant MCEs, as listed in Table II.

As another important measure of the MCE, the refrigerant capacity (RC) for different $\Delta H$s was also calculated by integrating the $\Delta S_m$ curve over the FWHM, as shown in Fig. 4(b). A linear relation is observed between RC and $\Delta H$, with a maximum of 162 J kg$^{-1}$ at $\Delta H$ of 50 kOe (Fig. 4(b) and Table II). Considering the magnetic hysteresis shown in isothermal M(H) curves (Fig. 3), the hysteresis loss (HL), with a maximum of 8 J kg$^{-1}$, for the field change of 50 kOe was further estimated by calculating the areas enclosed by the magnetization and demagnetization M(H) curves, as shown in inset to Fig. 4(b). A very small average HL of 5 J kg$^{-1}$ (Table II) was estimated over the same temperature range used for calculating the RC. Taking the average HL into account, a net RC of 157 J kg$^{-1}$ was subtracted (red diamond in Fig. 4(b); Table II). Although $\Delta S_m$ of -8 J K$^{-1}$ kg$^{-1}$ is relatively small, the reversible RC of 157 J kg$^{-1}$ is larger than those of many first-order magnetostructural transitions (Table II), owing to the successive magnetic phase transitions in this alloy.

To summarize, the martensitic and magnetic transitions of hexagonal MnNi$_{0.8}$Fe$_{0.2}$Ge alloy were studied. The results confirmed a coupling of first-order PM-to-induced-FM martensitic and second-order AFM-to-FM metamagnetic transitions in this alloy. Based on this coupling, MnNi$_{0.8}$Fe$_{0.2}$Ge alloy exhibits collective magnetic-related physical effects, including a magneto-volume effect and a table-like magnetocaloric effect. The magneto-volume effect is driven by the compressive stress produced in the volume-expansion martensitic transition, by which the important parameter of internal stress in the martensitic transition can be in reverse obtained. The table-like magnetocaloric effect based on two successive magnetic phase transitions shows a large net magnetic refrigerant capacity around room temperature.

This work is supported by the National Natural Science Foundation of China (51301195, 11174352, and 51171206) and the National Basic Research Program of China (973 Program: 2012CB619405).

TABLE I. Phase structures and lattice parameters at room temperature.

| Structure | $a_o\,(c_h)$ | $b_o\,(a_h)$ | $c_o\,(\sqrt{3}a_h)$ | $V_o\,(2V_h)$ |
|---|---|---|---|---|
| Hex. | 5.384 | 4.095 | 7.093 | 156.36 |
| Orth. | 6.032 | 3.781 | 7.097 | 161.87 |
| Δ, % | +12.04 | -4.41 | +0.06 | +3.52 |

Notes: 1) the lattice parameters are given in an orthorhombic description;

2) Δ = ($x_{orth.}$ - $x_{hex.}$)/$x_{hex}$, $x$ represents $a$, $b$, $c$ axes and $V$, respectively.

TABLE II. Magnetic-entropy change ($\Delta S_m$), working temperature range ($T_{hot}$ and $T_{cold}$: the extreme temperature ends of the full width at half maximum (FWHM) of the peak in $\Delta S_m$-T curve), refrigerant capacity (RC) and average hysteresis loss (HL) at 50 kOe of MnNi$_{0.8}$Fe$_{0.2}$Ge and various giant MCE materials.

| Compounds | $\Delta S_m$ (J kg$^{-1}$ K$^{-1}$) | Working temperature range (K) | | | Refrigerant capacity (J kg$^{-1}$) | | | Refs. |
|---|---|---|---|---|---|---|---|---|
| | | $T_{cold}$ | $T_{heat}$ | FWHM | RC | Average HL | Net RC | |
| MnNi$_{0.8}$Fe$_{0.2}$Ge | -8 | 284 | 310 | 26 | 162 | 5 | 157 | Present |
| MnNi$_{0.77}$Fe$_{0.23}$Ge | -19 | 260 | 272 | 12 | | | | 5 |
| Mn$_{0.9}$Co$_{0.1}$NiGe | -40 | 236 | 241 | 5 | 159 | | | 32 |
| Mn$_{0.965}$CoGe | -26 | 283 | 293 | 10 | | | | 21 |
| Mn$_{1.05}$Ni$_{0.85}$Ge | 27 | 132 | 140 | 8 | | | | 33 |
| Mn$_{0.89}$Cr$_{0.11}$NiGe | -28 | 269 | 279 | 10 | 236 | | | 34 |
| Mn$_{0.92}$Cu$_{0.08}$CoGe | -53.3 | 313 | 317 | 4 | | | | 35 |
| MnCoGeB$_{0.02}$ | -47.3 | 275 | 281 | 6 | | | | 36 |
| MnAs | -32 | 316 | 332 | 16 | | | | 37 |
| Ni$_{50}$Mn$_{37}$Sn$_{13}$ | 18 | 297 | 303 | 6 | | | | 15 |
| Ni$_{50}$Mn$_{34}$In$_{16}$ | 19 | 230 | 242 | 12 | 181 | 77 | 104 | 38 |
| Ni$_2$Mn$_{0.75}$Cu$_{0.25}$Ga | -65 | 307.4 | 309 | 1.6 | 84 | 12 | 72 | 18 |
| Ni$_{42}$Co$_8$Mn$_{30}$Fe$_2$Ga$_{18}$ | 31 | 203 | 208 | 5 | 110 | 40 | 70 | 39 |
| Ni$_{52}$Mn$_{26}$Ga$_{22}$ | -30 | 353 | 356 | 3 | 75 | 5 | 70 | 40 |
| Ni$_{46}$Co$_4$Mn$_{38}$Sb$_{12}$ | 32.3 | 291 | 295 | 4 | 95 | 21 | 74 | 41 |
| Fe$_{50}$Rh$_{50}$ | 16.4 | 385 | 400 | 15 | 201 | 53 | 148 | 42 |

**Fig. 1**. (Color online) Room-temperature XRD pattern, showing the coexistence of hexagonal austenite and orthorhombic martensite phases. Inset shows the magnetostructural phase diagram of MnNi$_{1-x}$Fe$_x$Ge. The data were taken from Ref. 5.

**Fig. 2**. (Color online) (a) M(T) curves in various magnetic fields. (b) Field-dependent magnetic transition temperatures in martensite phase. (c) Magnified image of M(T) curves from (a). Region 1 and 2 represent the PM-to-induced-FM MT and the AFM-to-FM metamagnetic transition, respectively. (d) Weak first-order AFM ordering transitions at $T_N^M$ indicated by cooling/heating M(T) curves. Inset illustrates the magneto-volume effect around $T_N^M$ driven by martensitic transition. Curves of MnNiGe and MnNiGe$_{0.98}$Sn$_{0.02}$ were replotted with data taken from Refs. 5, 26.

**Fig. 3**. (Color online) M(H) curves at various temperatures during the successive magnetic phase transitions. Region 1 and 2 represent the PM-to-induced-FM MT and the AFM-to-FM metamagnetic transition, respectively. Inset shows the spiral AFM-to-FM metamagnetic transition at 275 K.

**Fig. 4**. (Color online) (a) Isothermal magnetic-entropy changes ($\Delta S_m$) for various field changes ($\Delta H$). Inset shows the relations between $\Delta S_m$ and $\Delta H$ for the PM-to-induced-FM MT and the AFM-to-FM metamagnetic transition. (b) Refrigerant capacity (RC) for different $\Delta H$s. Inset shows the hysteresis loss (HL) for the $\Delta H$ of 50 kOe.